\documentclass[onecollarge]{svjour2}
\usepackage{amsmath,amssymb}
\usepackage[dvips]{graphicx}

\newcommand{\be}{\begin{eqnarray}}
\newcommand{\ee}{\end{eqnarray}}

\def\ben{\begin{equation}}
\def\een{\end{equation}}
\def\bena{\begin{eqnarray}}
\def\eena{\end{eqnarray}}

\begin{document}

\title{Matching WMAP 3-yrs results with the Cosmological Slingshot
  Primordial Spectrum}

\author{Cristiano Germani \and Michele Liguori}
\institute{Cristiano Germani \at SISSA and INFN, via Beirut 4, 34014
  Trieste, Italy,\\\email{germani@sissa.it}
  \and Michele Liguori \at D.A.M.T.P., Centre for Mathematical
  Sciences, University of Cambridge, Wilberforce
  road, Cambridge CB3 0WA, England,\\\email{M.Liguori@damtp.cam.ac.uk}}

\date{}
\maketitle

\begin{abstract}
We consider a recently proposed scenario for the
generation of primordial cosmological perturbations, the so called Cosmological
Slingshot scenario. We firstly obtain a general expression for
the Slingshot primordial power spectrum which extends previous
results by including a blue pre-bounce residual contribution at large scales.
Starting from this expression we numerically compute the CMB
temperature and polarization power spectra arising from the Slingshot
scenario and show that they excellently match the standard WMAP
3-years best-fit results. In particular, if the residual blue spectrum is far above the largest WMAP observed scale,
the Slingshot primordial spectrum fits the data well
by only fixing its amplitude and spectral index at the pivot scale $k_p=10^{-3}h\mbox{Mpc}^{-1}$.
We finally show that all possible distinctive
Slingshot signatures in the CMB power spectra are confined to very
low multipoles and thus very hard to detect due to large cosmic
variance dominated error bars at these scales.
\end{abstract}
%\begin{keyword}
% keywords here, in the form: keyword \sep keyword
% PACS codes here, in the form: \PACS code \sep code
%\PACS{11.25.Wx,98.80.-k}
%\end{keyword}

\begin{flushleft}
SISSA 37/2007/A; DAMTP-2007-50
\end{flushleft}

\section{Introduction}

\begin{figure}
\begin{center}
\includegraphics[width=0.8\textwidth, height=0.5\textheight]{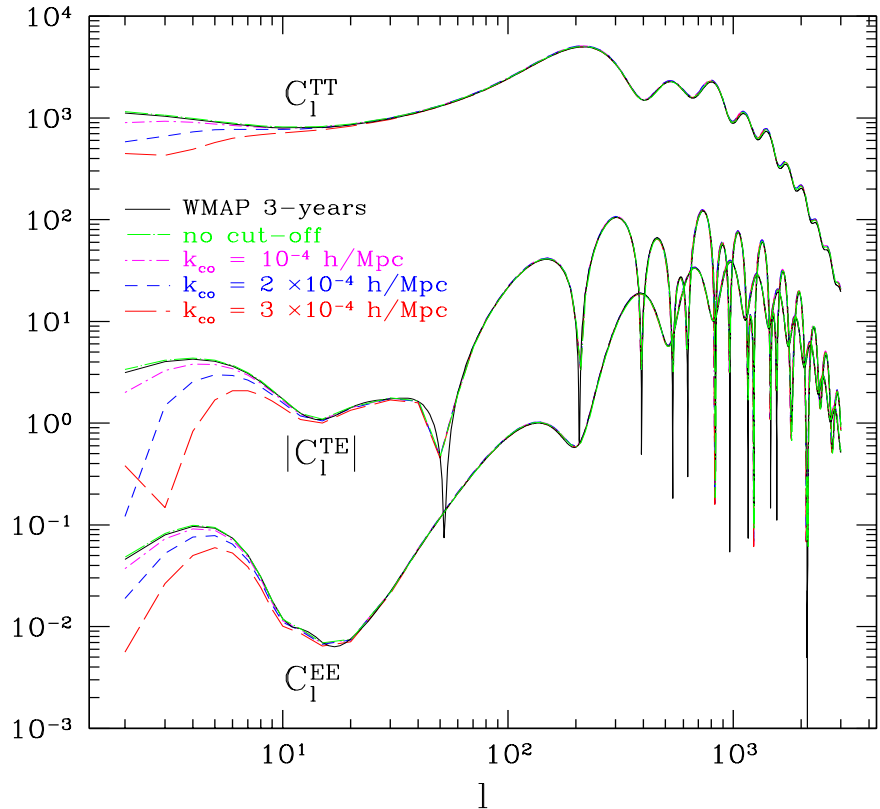}
\caption{Temperature and polarization CMB power spectra in the slingshot
  scenario, compared to the standard WMAP 3-years best-fit
  cosmological model. The cosmological parameters $\Omega_b h^2
  =0.02218$, $\Omega_c h^2 =0.1010$, $\Omega_\Lambda = 0.77$, $\tau =
  0.09$, $h = 0.74$ are the same in the two model. The primordial
  spectral index in the standard scenario has been chosen as $n =
  0.95$, while the slingshot primordial power spectrum is described in
  section II. The slingshot power spectrum normalization is
  chosen in order to match the amplitude of the CMB angular power
  spectrum at $\ell=100$ obtaining $k_0=8.86648\times 10^{-7}\ h\mbox{Mpc}^{-1}$.
  We consider several different cut-off scales and $B \ll \bar{B}$
 (see text for further details).}\label{fig:Cls}
\end{center}
\end{figure}

It is well known that Standard (non-inflationary) Cosmology is
afflicted by three severe problems \cite{KolbTurner}: homogeneity, isotropy and flatness.
Inflation is the standard accepted paradigm for the resolution of
these problems.
Nevertheless, as a fundamental origin of Inflation is as yet
lacking, many attempts to alternatively solve the homogeneity,
isotropy and flatness
fine tunings have been recently put forward.

In this paper we consider one of these alternatives, namely the
scenario developed
in \cite{sling,sling2} so called the
``Cosmological Slingshot Scenario'', or shortly the {\it Slingshot}.

In the Slingshot, our Universe is a probe $D3$-brane ``orbiting'' with an open trajectory
in a IIB supergravity background, namely the
Klebanov-Tseytlin (KT) metric \cite{KT} (the bulk).
If the probe brane approach of \cite{kehagias} used in \cite{sling,sling2} can be used,
the Slingshot trajectory results on an induced cosmological evolution on
the brane. More precisely,
a brane observer experiences a Friedman-Robertson-Walker non-singular
bouncing universe.
In the Slingshot, the problems that afflict standard cosmology are circumvented \cite{sling} by using similar mechanisms introduced in
pre big-bang \cite{veneziano} and cyclic \cite{turok} scenarios.
Besides, the Slingshot also predict a power spectrum of primordial perturbations. In
\cite{sling,sling2}, and in this letter, the
primordial spectrum of scalar perturbations due to the fluctuation of
the Slingshot brane on the KT background is indeed calculated under the
approximation that the backreaction of the Slingshot brane into the
bulk is negligible. The validity of this approximation is supported by the fact that the KT
background, in which the Slingshot brane is moving, is produced by a large number of D3-branes
having all the {\em same}
tension (``mass'') as the Slingshot brane.

The plan of the paper is as follows. In section \ref{sec:summary} we will
summarize the previous results of \cite{sling,sling2}. We
will then extend those results in two ways. First of
all we will consider a new blue contribution to the primordial power
spectrum that was previously not accounted for. This
will produce the general parametrization of the Slingshot primordial
spectrum shown at the end of section \ref{sec:blue}.

As a second step, in section \ref{sec:CMB}, we will use
this general parametrization in order to numerically
compute the temperature and polarization CMB angular power
spectra arising from the Slingshot and we will compare them to WMAP
data \cite{wmap}. In particular we will show that a suitable and
natural choice of the Slingshot parameters {\em allow to reproduce
the standard WMAP 3-years best-fit power spectra}. We will then try different choices
for the Slingshot parameters and see if they can produce distinctive model-dependent signatures
in the results. Finally we will draw our conclusions in section \ref{sec:conclusions}.

We are now ready to conclude this section, but we would like to make a final remark first.
In \cite{sling2} an analytic expansion of the Slingshot spectrum
for large multipoles ($\ell>10$) had actually already been shown to match
the WMAP best fit of a power law spectrum
with spectral index $n_s\simeq 0.95$. However this result held {\em only}
at a given {\it pivot scale} (chosen as $k_p\sim 10^{-3}h \mbox{Mpc}^{-1}$).
Our numerical approach in this paper shows instead that the spectrum found in \cite{sling2},
matches the WMAP experimental results at {\it all}
scales, and not only around the pivot scale.
This is not an obvious result as the Slingshot primordial spectrum presents a
non-trivial running of the spectral index.

%%%%%%%%%%%%%%%%%%%%%%%%%%%%%%%%%%%%%%%%%%%%%%%%%%%%%%%%%%%%%%%%%%%%%%%%
\section{The Original Slingshot Power Spectrum}\label{sec:summary}

\begin{figure}
\begin{center}
\includegraphics[width=0.8\textwidth, height=0.5\textheight]{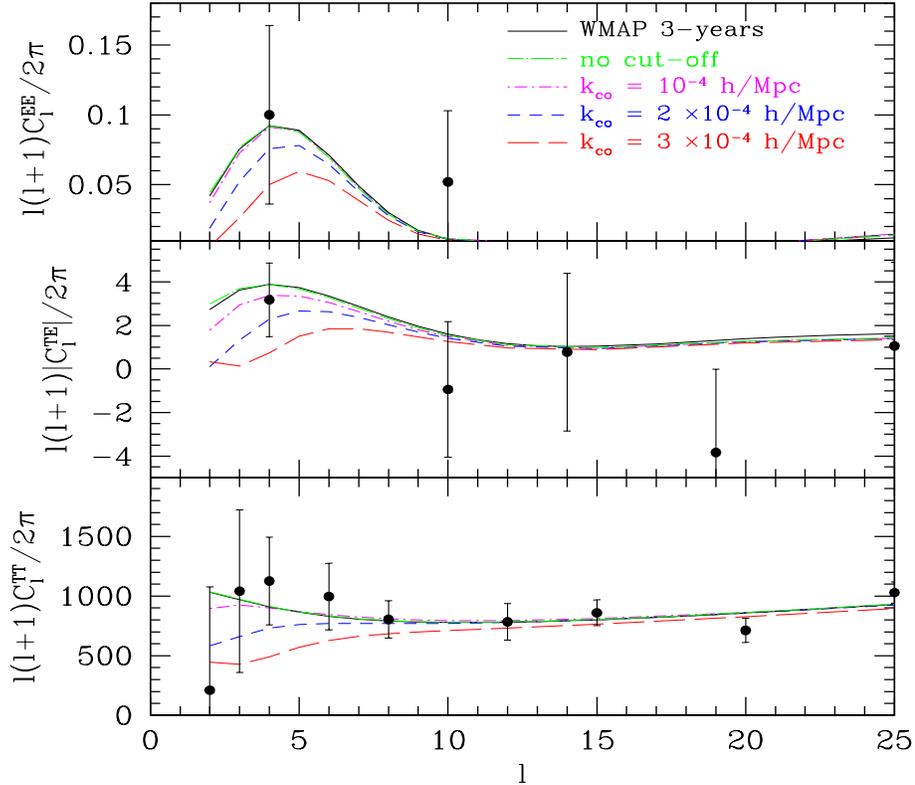}
\caption{Comparison between WMAP data and low CMB multipoles
for different values of $k_{cut-off}$ and fixed $B \ll \bar{B}$ in the Slingshot scenario.
}\label{fig:cutoff}
\end{center}
\end{figure}

The Slingshot power spectrum of scalar perturbations
is related to the quantum fluctuation of the Slingshot $D3$-brane. The way of
producing this perturbation is similar to the one introduced
by \cite{wald} but without the drawbacks
outlined by \cite{muk} (see \cite{sling2}). The fluctuation of the brane
is of quantum origin and it is in a pure
state whenever the comoving wave length of the perturbation is below
a fundamental quantum length $l_c$. This fundamental length can be consistently chosen
to be the first massive mode of the fundamental String or the M-theory minimal length.

During the pre-bounce phase of the Slingshot, the perturbations
created in the far past (in the Bunch-Davis vacuum), eventually come
back to their vacuum state whenever their wavelength $\lambda=a/k<l_c$,
where $k$ is the wave number of the fourier mode associated
with the wavelength $\lambda$ and $a$ is the scale factor of the
induced cosmology on the brane.
Viceversa, as the brane re-expands, the
wavelength of a given perturbation will grow again.
It is therefore clear that after some time, the perturbation
wavelength will reach the scale $l_c$.
At this point
the perturbation collapses into its classical state becoming
a coherent state as the perturbation is over-damped
by the Universe expansion \cite{Liddle}.
After waiting enough time, a stochastic background of primordial
perturbations is therefore dynamically (and continuously)
created. The distribution of these perturbations is
gaussian with variance set by the quantum correlations
during the quantum to classical transition.
However, since the Slingshot Scenario represent a bouncing cosmology,
not all the wavelength of primordial perturbations
can be produced with this mechanism. In fact the maximal wavelength that can be produced is
\be
\lambda_{\mbox{\tiny cut-off}}=\frac{a_b}{k_{\mbox{\tiny cut-off}}}=l_c\ ,
\ee
where $a_b$ is the size of the scale factor at the bouncing.
This obviously creates a natural cut-off on the power
spectrum.

A remark here is due. The physical process we have just discussed has been developed in the String frame. There,
at zeroth order on the brane velocities \cite{sling2}, the
gravitational coupling is running ($G_N\sim a^2$) and particle masses are fixed.
The Einstein frame, in which the the gravitational coupling is constant, can be then easily obtained by re-scaling the metric by $a^{-2}$.
At the background level then, the spacetime in Einstein frame is Minkowski and all particle masses run
(note that also the matter Lagrangian is re-scaled).
It is therefore easy to convince ourselves that any physical quantity in the two frames is completely equivalent
(see \cite{faraoni} for a general discussion and \cite{sling2} for the Slingshot case).
Let us however comment the special case of the perturbation spectrum.
In String frame the perturbed metric is $ds^2_S=a^2(1+2\Psi)dt^2+...$ where $\Psi$
is the Bardeen potential. By re-scaling to the Einstein frame such that $ds^2_E=(1+2\Psi)dt^2+...$,
we still obtain the same Bardeen potential $\Psi$. The power spectrum of primordial perturbations (a physical quantity) is therefore unchanged
by the change of conformal frames. However, one might still be puzzled whether the String frame cut-off on the perturbations spectrum is still
there in the Einstein frame.
In the Einstein frame the wavelength of a perturbation is constant in time, {\it i.e.} $\lambda=1/k$. However,
the quantum length in this frame is re-scaled, together with any other physical length, by a factor $a^{-1}$.
In the Einstein frame the quantum length is therefore ``bouncing''.
This produces, in the Einstein frame, the same cut-off as observed in the String frame.

In \cite{sling2}, the Slingshot primordial power spectrum is calculated to be
\be\label{eqn:Pkslingshot}
P_{\mbox {\tiny
    red}}(k)=\frac{A}{k}e^{\frac{1}{2}W_{-1}\left(-\frac{k_0^4}{k^4}\right)}
\left[1-e^{\frac{1}{2}\Delta(k_0/k;k_{\mbox{\tiny cut-off}})}\right]\ ,
\ee
where $W_{-1}$ is the Lambert W function in the real branch $-1$ (see \cite{WW}
for a description of the Lambert function properties), $A$
is an overall normalization of the spectrum and
finally $k_0$ is a parameter defining the spectral index at large $k$s.

The cut-off function $\Delta(k_0/k;k_{\mbox{\tiny cut-off}})$ is defined as
\be
\Delta(k_0/k;k_{\mbox{\tiny cut-off}})=W_{-1}\left(-\frac{k_0^4}{k^4}\right)-
W_{-1}\left(-\frac{k_0^4}{k_{\mbox{\tiny cut-off}}^4}\right)\ ,
\ee
where $k_{\mbox{\tiny cut-off}}$ fixes the cut off as $P_k$
is positive definite. If the ratio $k_0/k$ is small we have
$W_{-1}(-k_0^4/k^4)\simeq 4\ln (k_0/k)$. Therefore for large wave number ({\it i.e.} small scales)
\be
e^{\frac{1}{2}W_{-1}\left(-\frac{k_0^4}{k^4}\right)}\simeq \frac{k_0^2}{k^2}\ .
\ee
In this limit $e^{\frac{1}{2}\Delta}\ll 1$ and the spectrum looks scale invariant.

In order to match the best WMAP fit of a power law power spectrum,
following \cite{sling2}, we fixed the spectral index of the Slingshot to
be $n_s=d\ln k^3P(k)/d\ln k+1=0.95$ at the pivot scale $\ell=100$.
If we neglect the correction due to the cut-off scale,
this fixes the parameter $k_0$ to be $k_0=8.86648\times 10^{-7}\ h\mbox {Mpc}^{-1}$.
This parameter is much smaller than any wave number we are
going to consider. We can then use the analytical properties of the
Lambert W function $W_{-1}(-x)\simeq \ln(x)-\ln(-\ln(x))$, to find the approximate
spectrum
\be\label{eqn:Pkapprox}
P_{\mbox {\tiny red}}(k)\simeq \frac{A}{k^3\sqrt{\ln\frac{k}{k_0}}}
\left[1-\frac{k_{\mbox{\tiny cut-off}}^2}{k^2}
\sqrt{\frac{\ln \frac{k_{\mbox{\tiny cut-off}}}{k_0}}{\ln
\frac{k}{k_0}}}\right]\ .
\ee
This completes the description of the previous results obtained in \cite{sling,sling2}.
In the following section we are going to consider an additional contribution to the Slingshot
primordial power spectrum that was not kept into account in previous works.

%%%%%%%%%%%%%%%%%%%%%%%%%%%%%%%%%%%%%%%%%%%%%%%%%%%%%%%%%%%%%%%
\section{A blue pre-bounce residual}\label{sec:blue}

\begin{figure}
\begin{center}
\includegraphics[width=0.9\textwidth, height=0.5\textheight]{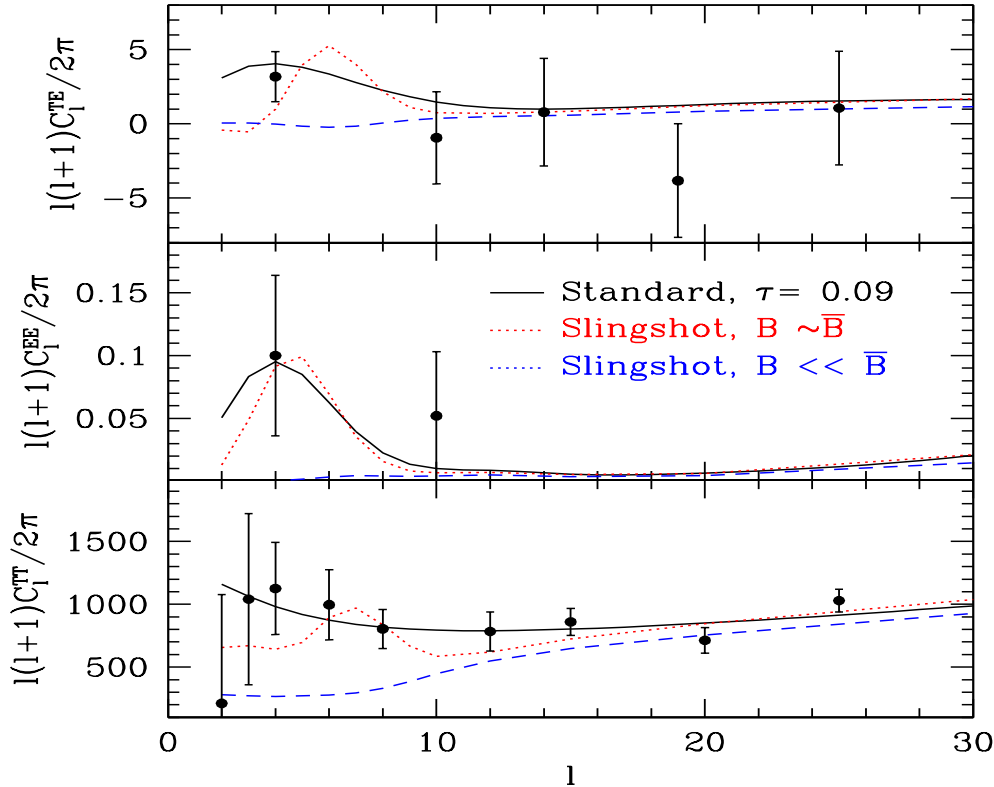}
\caption{We compare different values of the low-$\ell$ spectrum
  normalization $B$ for a given value of $\ell_{cut-off} \sim 10$ and
  optical depth to last scattering $\tau = 0.09$. The solid black line represents
  the standard WMAP 3-years best-fit power spectrum while the dashed
  blue and dotted redlines are the Slingshot power spectra for small
  and large values of $B$ respectively.}\label{fig:tau009}
\end{center}
\end{figure}

As previously discussed, primordial perturbations are in general
present even
during the pre-bounce era. These perturbations, corresponds to the
quantum fluctuation of the Slingshot brane during its motion down the
throat of the CY.
The induced Bardeen potential $\Phi$ evolves
with the angular brane motion and its perturbations, as discussed in
\cite{sling2}.
Nevertheless, we can use the approximations used in \cite{sling} where
the Bardeen potential results decoupled from
the angular brane motion.

In this case, $\Phi$, follows schematically the Mukhanov equation
\cite{sling}
\be
\delta r''+\left(k^2-\frac{J^2}{r^4}\right)\delta r=0\ ,
\ee
where $r\Phi=\delta r$, $J$ is the brane angular momentum, $r$ parameterize
the brane position in the CY throat and finally $'$ is the derivative with
respect to the conformal time. $J^2/r^4$ corresponds to $r''/r$. Differently
from the inflationary case, however, $r''/r$ {\it is not} the Hubble
horizon.

The induced scale factor of the Universe $a$ is related to $r$ as
$a=\frac{r}{L\sqrt{\ln r/r_s}}$.
$L$ is proportional to the number $N$ of $D3$-branes in the stack and
$r_s$ is the radius of the blown up sphere at the tip of the CY (see
\cite{sling} for more details).
We then see that for $k\gg \frac{J}{r^2}$ the system oscillates,
particles are not created and the Bardeen
potential stays in its vacuum.
In the opposite case, $k\ll \frac{J}{r^2}$,
the system is instead over-damped and
eventually $\Phi$ becomes constant. There, particles are created and
the system evolves stochastically. In the large $k$ region
we therefore have
$\langle\Phi(k)\Phi(k')\rangle\propto \frac{\delta(k,k')}{k r^2}$.
At the matching point $k=\frac{J}{r^2}$, we then have a constant
spectrum of perturbations, {\it i.e.}
a power law spectrum with spectral index $n_s=4$.
Note that if $J=0$, {\it i.e.} for a brane with no-angular momentum,
the perturbation is never over-damped and therefore the spectrum will
be $P(k)\propto k^{-1} r^2$ at any times. In this case
the resulting spectral index will be $n_s=3$ as found in \cite{bran}.

These conclusions can also be drawn more precisely by following
\cite{sling2} and by considering the semi-classical to quantum matching point at
$k=Jr^{-2}$.

We then conclude that a blue spectrum of primordial perturbations,
coming from the pre-bounce, must be added to the $P_{\mbox{\tiny red}}(k)$.
However, this spectrum will survive from being destroyed by the
quantum region only for perturbation scales $k<k_{\mbox{\tiny cut-off}}$, as discussed
before. Therefore the full spectrum of perturbation turn out to be
\begin{equation}\label{eqn:Pk}
P(k) = \left\{ \begin{array}{ll}
       P_{\mbox{\tiny red}}(k),&
       \textrm{if $k > k_{\mbox{\tiny cut-off}}$} \\
       P_{\tiny blue}(k) \equiv \frac{B}{k_0^3} & \textrm{if $k < k_{\mbox{\tiny cut-off}}$\ .}
       \end{array} \right. \; ,
\end{equation}
where the amplitude $B$ is a completely free parameter.
The blue part of the spectrum (\ref{eqn:Pk}) did not appear previously
in the literature. Thus, eq. (\ref{eqn:Pk})
constitutes the most general parameterizations of the Slingshot power spectrum and
completes the previous results of \cite{sling, sling2}. In the following section
we will numerically compute the CMB angular power spectra arising from this primordial spectrum.

\section{Matching the WMAP results}\label{sec:CMB}

\begin{figure}
\begin{center}
\includegraphics[width=0.9\textwidth, height=0.5\textheight]{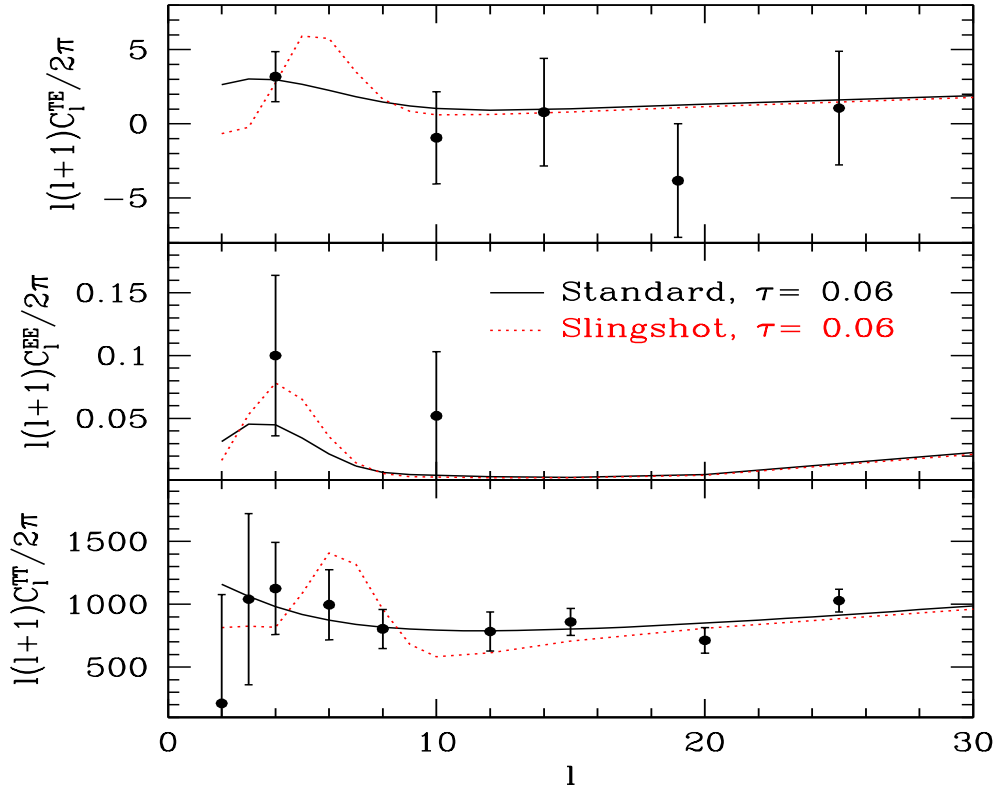}
\caption{We consider a relatively low value of the optical depth to
  reionization, $\tau=0.06$, and study the effect of varying the
  normalization parameter $B$ in order to look for possible
  degeneracies between $\tau$ and $B$. The solid black line is the
  standard inflationary power spectrum whereas the dotted red line is
  the Slingshot prediction for a suitably large value of $B$. Even if
  we can improve the fit of polarization data with respect to the
  standard case, we produce a bump in low-$\ell$ temperature spectrum
  which does not allow to fit the data well
  ($\Delta \chi^2 =1219$)
.}\label{fig:tau006}

\end{center}
\end{figure}

The CMB temperature and polarization angular power spectra are
obtained from the primordial power spectrum of scalar perturbations
$P(k)$ through the well-known formula (see e.g. \cite{Seljak}):

\ben\label{eqn:Pk2Cl}
C^{XX}_\ell = (4 \pi)^2 \int_{k_{min}}^{k_{max}} dk k^2 P(k) (\Delta^X_\ell(k))^2 \; ,
\een

where $\Delta^X_\ell(k)$ are the radiation transfer functions and $X=T,E$
defines temperature and polarization respectively. The temperature and
polarization transfer functions can be extracted from a Boltzmann code
like e.g. CMBfast. The angular power spectrum predicted by the
Slingshot scenario can then be calculated starting from the primordial
power spectrum of formula (\ref{eqn:Pk}) and numerically evaluating the
Lambert W-functions. The Slingshot power spectrum
expression contains two new free parameters: the cut-off scale
$k_{cut-off}$ and the amplitude $B$ of the large scale part $P_{blue}(k)$.
We will now study the effects of varying these parameters on the final $C_\ell$s.

The first alternative we consider is to choose $k_{cut-off}$
such that $k_{cut-off} \ll k_{min}$ where $k_{min}$
is the smallest wavenumber appearing in the integral defined by
eqn. (\ref{eqn:Pk2Cl}).
A cut-off below $k_{min}$ then clearly  affects only scales that are
unobservable. The choice of $B$ is then not relevant in this case
and we can replace $P(k)$ in eqn. (\ref{eqn:Pk2Cl}) with $P_{red}(k)$
defined in eqn. (\ref{eqn:Pkslingshot}). We calculated the CMB power
spectrum using this power spectrum and transfer functions obtained
from the WMAP 3-years best-fit cosmological parameters. The resultant Slingshot
$C_\ell$s in this case are represented by the dot-dashed green line
in Figure \ref{fig:Cls}. In the same Figure, the solid black line
represents the standard WMAP best-fit power spectrum. The two spectra
present a very good match, thus showing that,
just by fitting the two parameters $A$ and $k_0$, {\em the Slingshot
model allows to well reproduce the standard CMB angular power
spectrum from WMAP}. More precisely, an explicit calculation of the likelihood
for the Slingshot shows that the goodness-of-fit relative to
the standard WMAP 3-years spectrum is $\Delta \chi^2_{eff} = 3$, which, being non statistically significant \cite{wmap},
makes the slingshot still a good fit of the data.
As we were already stressing
above, this result was not obvious due to the non-trivial running of
the Slingshot primordial power spectrum.

Let us now consider a cut-off on scales that are relevant for the
CMB, i.e. $k_{cut-off} \gtrsim k_{min}$ in eqn. \ref{eqn:Pk2Cl}.

In this case the wavenumber $k_{cut-off}$ will roughly define an angular cut-off
$\ell_{cut-off}$ below which the angular power spectrum is basically obtained
from a constant primordial power spectrum $P(k) = P_{\tiny blue} \equiv
B/k_0^3$ (see eqn. (\ref{eqn:Pk})).
At this point we found it useful for our analysis
to determine a value of the normalization parameter $B$ which makes
the amplitudes of $P_{\tiny red}(k)$ and $P_{\tiny blue}(k)$ to
roughly coincide at the pivot scale (that we chose to be $k \sim
10^{-3}$ in our analysis). In order to match WMAP data for $\ell >
\ell_{cut-off}$ we need $|k^3 P_{\tiny red}(k)| \sim 10^{-10}$.
Thus matching the two amplitudes yields:

\ben
k_{p}^3 \frac{B}{k_0^3} = k_{p}^3  P_{\tiny red}(k_{p})
\sim 10^{-10} \; .
\een

With our values of $k_0 \sim 10^{-6}$ and $k_{p} \sim 10^{-3}$, we get
$B \equiv \bar{B} \sim 10^{-19}$. We can now distinguish between two
cases: $B \ll \bar{B}$ and $B \gtrsim \bar{B}$.

Let us firstly take
$B \ll \bar{B}$ and consider several different $k_{cut-off}$. In this case
the amplitude of $P_{\tiny blue}$ is much smaller than the amplitude
of $P_{\tiny red}$. We then expect to see a suppression of power on scales
$\ell < \ell_{cut-off}$. This is shown in figures \ref{fig:Cls} and
\ref{fig:cutoff}. The same
pictures also suggest that a cut-off scale $k_{\mbox{\tiny cut-off}}
\lesssim 2 \times 10^{-4}$
is still a good-fit to the data: the goodness-of-fit relative to the
WMAP best-fit spectrum is $\Delta \chi^2_{eff} \leq 3$ in this range (for a similar discussion applied to a different model
see \cite{michele}).

Let's now move to the case $B \gtrsim \bar{B}$. We can now choose $B$ large
enough to eliminate the suppression of the larger angular scales
that we have just described above. In Figure
\ref{fig:tau009} we consider an angular cut-off scale
$l_{cut-off} \sim 10$ and we show that a choice of the normalization
$B \sim \bar{B}$ can significantly improve
the goodness-of-fit relative to the case $B \ll \bar{B}$ with the same
cut-off ($\Delta \chi^2_{eff} = -112$).
In other words this suggests that a full likelihood
analysis of the slingshot parameters (which is beyond the purpose of this work)
would show some degeneracy
between $k_{cut-off}$ and $B$. Nevertheless it is important to note that this
does not allow to arbitrarily
increase the cut-off scale. The slope of the angular power spectrum for
$\ell < \ell_{cut-off}$ is indeed completely different in the two regimes
 and this becomes rapidly evident for large
$\ell_{cut-off}$. We then conclude that also in this case any possible Slingshot-related
signature is unfortunately confined to the first few CMB
multipoles, characterized by a large cosmic variance. For this reason
it seems impossible to use the CMB $TT$, $TE$, and $EE$ angular power
spectra as a way to
discriminate between Slingshot and standard inflationary cosmology.
To this purpose further investigation in other directions might be interesting
(e.g. non-Gaussian signatures, gravitational wave background).

Even if we give up the idea of finding
specific observable signatures of
the Slingshot model in the CMB temperature and polarization power
spectra, we are still left with the interesting following question:
if we  {\em assume} the Slingshot as the scenario for the
generation of primordial fluctuations, and we repeat the analysis of
WMAP results in this framework, are we going to see any change in the
final cosmological parameters? In the
acoustic peaks region both the Slingshot and the inflationary power
spectrum have the same slope, so we already know that the answer to
the previous question is `no' for most of the parameters. As the only
differences can be at small $\ell$, it seems
that the only parameter that can in principle be affected is the
optical depth at reionization $\tau$. Let us elaborate on this. WMAP
is known to predict a large optical depth to reionization $\tau \simeq
0.09$, or equivalently an early reionization at a redshift $z \sim 10$.
The signature of this early
reionization is mainly in the bump observed at low $\ell$ in the TE
and EE angular power spectra: there wouldn't be any primordial polarization
signal on large angular scales in absence of early reionization. This
conclusion still holds in the Slingshot scenario (it is only related to
the physics of Compton scattering). However the low-$\ell$ part of the
polarization spectrum is described in the Slingshot by three parameters, $B$,
$k_{cut-off}$ and $\tau$. Possible degeneracies among these parameters
could then eventually change the best-fit value of $\tau$.

In particular it might now happen
that a value of $\tau$ significantly smaller than in the standard scenario
could still allow a good fit of the low-$\ell$ polarization bump if we
compensate for it by
{\em increasing the amplitude} $B$. This also works in the opposite direction:
we can increase $\tau$ and reduce
$B$ accordingly. This mechanism is clearly very efficient if we
limit ourselves to considering polarization data only.
The situation however drastically changes
when we account for temperature data. A large $\tau$ produces a
low-$\ell$ bump in the TE and EE but it does not affect
the TT power spectrum. A large $B$ instead produces a large
enhancement of the low-$\ell$ TT power spectrum as well. This effect is not
compatible with the data if we have to compensate for a very
small (large) $\tau$ with a very large (small) $B$. In other words
temperature data contribute to largely breaking the degeneracy between
$\tau$ and $B$ that arises from polarization data alone. An example of
this is in Figure \ref{fig:tau006} where we try to fit
data with an optical depth $\tau = 0.06$ and all the other WMAP
parameters unchanged.

\section{Conclusions}\label{sec:conclusions}

In this paper we have studied some phenomenological implications of the
cosmological Slingshot scenario introduced in \cite{sling,sling2}. In
the first part of the paper we have derived an expression of the
primordial power spectrum of cosmological perturbations arising from
the Slingshot (formula \ref{eqn:Pk}). This expression generalizes
previous results by \cite{sling,sling2} in that it contains a blue
contribution to the spectrum which had not been considered before.
In the second part of the paper we have numerically computed
the CMB temperature and
polarization power spectra arising from the Slingshot primordial
spectrum. Firstly we showed that a suitable choice of the Slingshot
parameters allows to match Slingshot predictions with the WMAP 3-years best-fit
power spectrum. More precisely we showed through a relative
goodness-of-fit approach that the slingshot
predictions are not, in a statistical sense,
worse than the best WMAP fit of a power law primordial spectrum. This
conclusion has been drown by fitting the slingshot power spectrum spectral
index to be 0.95 (best WMAP fit) at some pivot scale and by keeping all the
standard cosmological parameter unchanged. To gain a more precise insight it
would be however very important to perform a Montecarlo analysis were all
parameter, in particular the spectral index, can change. This might in
principle find
a better fit to the data. However, as the aim of the present paper
was only to show that, with the same WMAP parameters, the slingshot power
spectrum is a good fit of the data, the above mentioned complete statistical
analysis is left for future work.
In the last part of the paper we finally looked for possible specific signatures of the
Slingshot in CMB data that could allow to distinguish it from the
standard scenario. Unfortunately all the distinctive Slingshot
features turn out to be confined to the low-$\ell$ part of the
spectrum, where large cosmic variance dominated error bars prevent
from any significant discrimination between the two scenarios.

\section*{Acknowledgments} We acknowledge the use of the Legacy Archive for
Microwave Background Data Analysis (LAMBDA). Support for LAMBDA is
provided by the NASA Office of Space Science. CG wishes to thank Pierstefano Corasaniti, Nicolas
Grandi and Alex Kehagias for useful discussions. CG wishes also to thank the Physics Department of King's
College London for the hospitality during part of this work.

\end{document}